\documentclass[multphys,vecarrow]{svmult}

\usepackage{graphicx}
\usepackage{amssymb}
\usepackage{amsmath}
\usepackage{subeqnar}
\usepackage{epsfig}
\usepackage{citesort}

\begin{document}
\frontmatter
\mainmatter

\bibliographystyle{unsrt}

\title*{Towards Noised-Induced Phase Transitions in Systems of Elements with
        Motivated Behavior}

\author{
 Ihor Lubashevsky\inst{1},
 Morteza Hajimahmoodzadeh\inst{2},
 Albert Katsnelson\inst{2}, and
 Peter Wagner\inst{3}
}
\institute{General Physics Institute, Russian Academy of Sciences,\\
           Vavilov str., 38, 119991 Moscow, Russia\\
           Contact address, e-mail: {\tt ialub@fpl.gpi.ru}\\
 \and
          Faculty of Physics, M.~V.~Lomonosov Moscow State University,\\
          Moscow 119992, Russia\\
          Contact address, e-mail: {\tt albert@solst.phys.msu.su}\\
 \and
    Institute of Transport Research, German Aerospace Center (DLR),\\
    Rutherfordstrasse 2, 12489 Berlin, Germany\\
    Contact address, e-mail: {\tt peter.wagner@dlr.de}\\
}
\titlerunning{Phase Transitions in Systems with Motivation}
\authorrunning{I. Lubashevsky, M. Hajimahmoodzadeh, A. Katsnelson, and P. Wagner}
\date{\today}
\maketitle

\begin{abstract}
A new type of noised-induced phase transitions that should occur in systems of
elements with motivated behavior is considered. By way of an example, a simple
oscillatory system $\{x,v=\dot{x}\}$ with additive white noise is analyzed
numerically. A chain of such oscillators is also studied in brief.
\end{abstract}

\section{Introduction}

Systems of elements with motivated behavior (systems with motivation), e.g.,
fish and bird swarms, car ensembles on highways, stock markets, \textit{etc.}
often display noise-induced phase transitions (for a review see
Ref.~\cite{Hel1}). The ability of noise to induce phase transitions is now well
established (see, e.g., Refs~\cite{In1,In2}). However, in systems with
motivation there is a special mechanism endowing the corresponding
noise-induced phase transitions with distinctive properties.

For example, people as elements of a certain system cannot individually control
all the governing parameters. Therefore one chooses a few crucial parameters
and focuses on them the main attention. When the equilibrium with respect to
these crucial parameters is attained the human activity slows down retarding,
in turn, the system dynamics as a whole. For example, in driving a car the
control over the relative velocity $v$ is of prime importance in comparison
with the correction of the headway distance $x$. So, under normal conditions a
driver should eliminate the relative velocity between her car and a car ahead
first and only then correct the headway.

These speculations lead us to the concept of dynamical traps, a certain ``low''
dimensional region in the phase space where the main kinetic coefficients
specifying the characteristic time scales of the system dynamics become
sufficiently small in comparison with their values outside the trap region
\cite{we2,we1}. The present paper analyzes the effect of noise on such a system
and demonstrates that additive noise in a system with dynamical traps is able
to give rise to new phases.

\section{Noised oscillatory system with dynamical traps}

By way of example, the following dimensionless system typically used to
describe the oscillatory dynamics is considered:
\begin{equation}
   \label{1.1}
   \frac{dx}{dt} =  v\,,\quad
   \frac{dv}{dt} = -\Omega (v)\left[x+ \sigma v\right] +\epsilon \xi(t)\,,
\end{equation}
where $\sigma$ is the damping decrement and the term $\epsilon\xi(t)$ is a
random Langevin ``force'' of intensity $\epsilon$ proportional to the white
noise $\xi(t)$ with unit amplitude. The function $\Omega (v)$ describes the
dynamical trap effect arising in the vicinity of $v=0$. For this function, the
following simple \textit{Ansatz}
\begin{equation}
 \label{1.3}
 \Omega (v)=\frac{v^{2}+\triangle ^{2}}{v^{2}+1}\,.
\end{equation}
is used. In the chosen scales the thickness of the trap region is equal to
unity and the parameter $\triangle \leq 1$ measures the trapping efficacy. When
$\triangle =1$ the dynamical trap effect is ignorable, for $\triangle =0 $ it
is most effective.

\begin{figure}[t]
\begin{center}
\includegraphics[width=110mm]{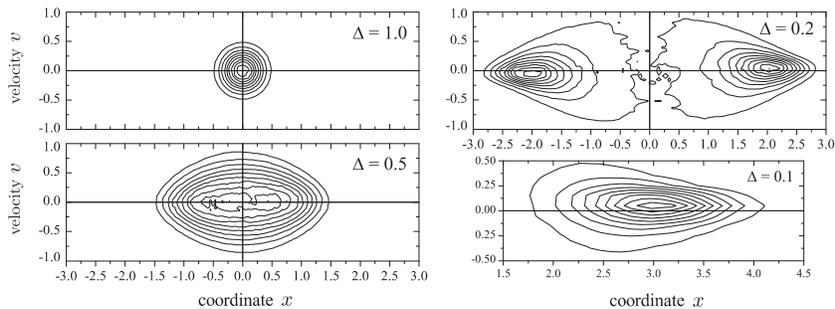}
\end{center}
\caption{Evolution of the distribution function $\mathcal{P}(x,v)$ (shown by
level contours) as the parameter $\triangle$ decreases. The parameters used are
$\sigma =0.1$ and $\epsilon = 0.1$. The lower right window depicts only one
maximum of the distribution function. \label{Fig2}}
\end{figure}

System~(\ref{1.1}), (\ref{1.3}) was analyzed numerically using the algorithm
described in \cite{SRK}. Figure~\ref{Fig2} shows the distribution function
$\mathcal{P}(x,v)$ of the system on the phase plane $\{x,v\}$, depending on the
parameter $\triangle$. As can be seen this system undergoes a second order
phase transition manifesting itself in the change of the shape of the phase
space density $\mathcal{P}(x,v)$ from unimodal to bimodal as the trap parameter
$\triangle$ decreases.

\subsection*{Mechanism of the phase transition}

\begin{figure}[t]
\begin{center}
\includegraphics[scale=0.85]{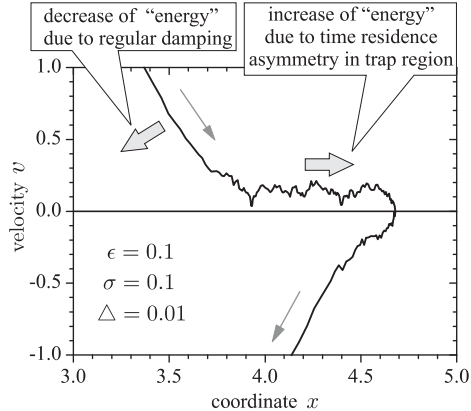}
\end{center}
\caption{A typical fragment of the system path going through the
   trap region. The parameters $\sigma =0.1$, $\epsilon =0.1$, and
   $\triangle =0.01$ were used in numerical simulations in order to
   make the trap effect more pronounced.\label{Fig4}}
\end{figure}

Figure~\ref{Fig4} illustrates the mechanism of this phase transition. It
depicts a typical fragment of the system motion through the trap region for
$\triangle \ll 1$. When the path goes into the trap region $\mathcal{Q} _{t}$
($|v|\ll 1$) the regular ``force'' $\Omega(v)(x +\sigma v)$  is depressed. So
inside this region the system dynamics is mainly random due to the remaining
weak Langevin ``force'' $\epsilon \xi (t)$. Crucial is the fact that the
boundaries $\partial_{+}\mathcal{Q}_{t}$ (where $v\sim 1$) and $\partial
_{-}\mathcal{Q}_{t}$ (where $v\sim -1$) are not identical in properties with
respect to the system motion. At the boundary $\partial _{+}\mathcal{Q}_{t}$
the regular ``force'' leads the system inwards the trap region
$\mathcal{Q}_{t}$, whereas at the boundary $\partial _{-}\mathcal{Q}_{t}$ it
causes the system to leave the region $\mathcal{Q}_{t}$. Outside the trap
region $\mathcal{Q}_{t}$ the regular ``force'' is dominant. Thereby, from the
standpoint of the system motion inside the region $\mathcal{Q}_{t}$, the
boundary $\partial _{+}\mathcal{Q}_{t}$ is ``reflecting'' whereas the boundary
$\partial _{-}\mathcal{Q} _{t}$ is ``absorbing''.

As a result the distribution of the residence time at different points of the
region $\mathcal{Q}_{t}$ should be asymmetric as shown in Fig.~\ref{Fig2} (it
is most clear in the lower right window). Therefore, during location inside the
trap region the mean velocity must be positive, causing the system to go away
from the origin. Outside the trap region the system motion is damping. So, when
the former effect becomes sufficiently strong the distribution function
$\mathcal{P}(\eta ,u)$ becomes bimodal.

\section{Chain of oscillators}
A similar noise-induced phase transition for an ensemble of oscillators with
dynamical traps is analyzed. The following one-dimensional model is considered,
$N$ balls can move along $x$-axis interacting with the nearest neighbors, which
is described by the equations for $i = 1,2,\ldots N-1$
\begin{equation}\label{s1}
   \frac{dx_i}{dt} =  v_i\,,\quad
   \frac{dv_i}{dt} = -\Omega_s (v_{i-1},v_{i},v_{i+1})\left[\eta_i+
   \sigma \vartheta_i\right] +\epsilon \xi_i(t)\,.
\end{equation}
Here $x_i$ is the coordinate of ball $i$, $v_i$ is its velocity, the variables
$\eta_i$ and $\vartheta_i$ are given by the expression:
\begin{equation}\label{a1}
    \eta_i = x_i -\tfrac12(x_{i-1} + x_{i+1})\,,\quad
    \vartheta_i = v_i -\tfrac12(v_{i-1} + v_{i+1})\,,
\end{equation}
and $\{\xi_i(t)\}$ is the collection of white noise sources of unit amplitude
and being mutually independent. The damping decrement $\sigma$ and the noise
intensity $\epsilon$ are assumed to be the same for all the oscillators. The
boundary balls ($i = 0$ and $i=N$) are let to be fixed to prevent the ball
system from moving as a whole. The function $\Omega_s (v_{i-1},v_{i},v_{i+1})$
measuring the trap effect due to the nearest neighbor interaction is given by
the \textit{Ansatz}
\begin{equation}\label{a2}
   \Omega_s (v_{i-1},v_{i},v_{i+1}) =\frac{(v_{i-1}-v_i)^2+(v_{i}-v_{i+1})^2
   +\triangle^2}
   {(v_{i-1}-v_i)^2+(v_{i}-v_{i+1})^2+1}\,,
\end{equation}
where the parameter $\triangle$ has the same meaning as previously, it measures
the intensity of trapping. The balls are assumed to be either mutually
permeable or impermeable. In the latter case the absolutely elastic collision
approximation is used.

The system of equations~(\ref{s1})--(\ref{a2}) was analyzed numerically. Below,
the results are presented for the impermeable balls. Again integration of the
stochastic differential equations was performed with the algorithms described
in \cite{SRK}. We analyzed an ensemble of 500 balls initially spaced 5 units
apart and counted the number of balls falling into 100 identical intervals
covering the system location region in the corresponding space.

\begin{figure}
\begin{center}
  \includegraphics[width=55mm]{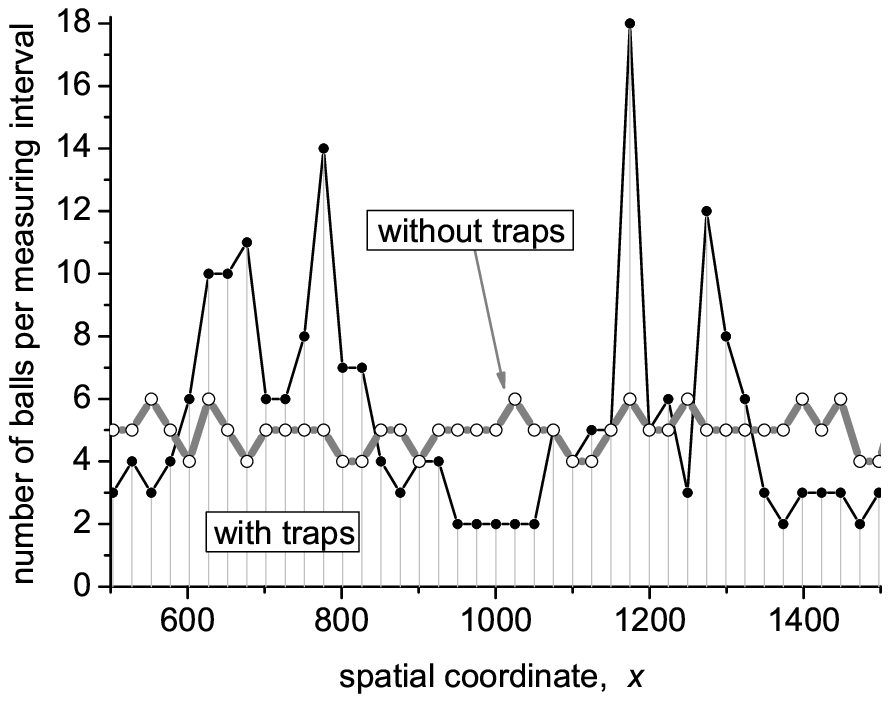}\quad
  \includegraphics[width=55mm]{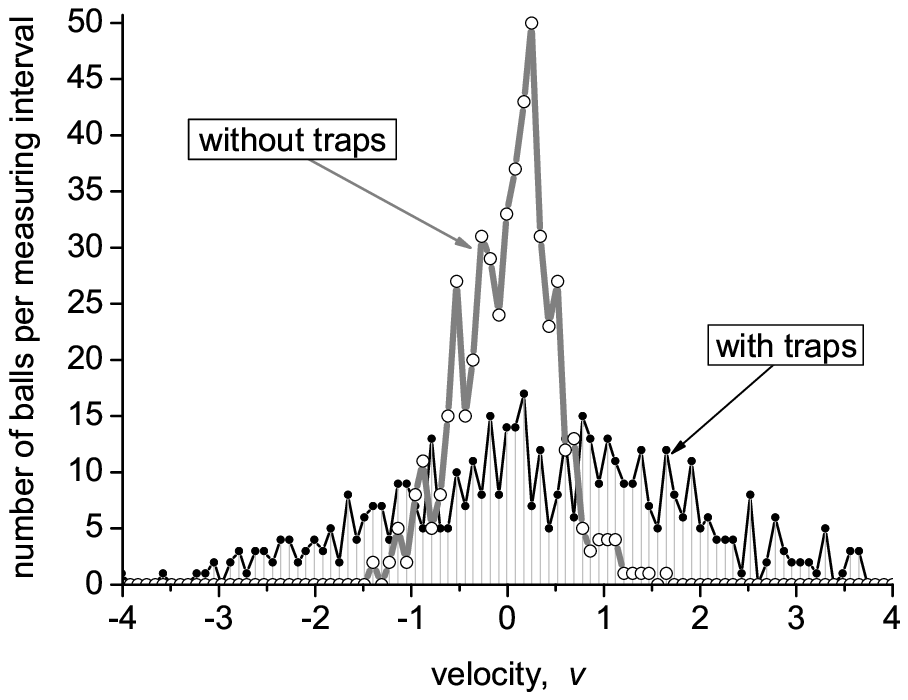}
\end{center}
  \caption{The ball distribution along the spatial axis and the velocity
  distribution at a fixed moment of time. These distributions are represented
  in terms of number of balls falling into measuring interval. The measuring
  intervals of number 100 cover the whole region of the system location in the
  corresponding space. The thick grey line and the boundary of dashed region
  match the cases where the trap effect is absent, $\triangle=1$,
  and strong, $\triangle=0.1$, respectively. The other parameters used are
  $\sigma =0.1$ and $\epsilon = 0.1$.}\label{FS1}
\end{figure}

Figure~\ref{FS1} actually depicts the ball distribution along the spatial axis
and the velocity space at a fixed moment of time. As seen, the trap effect
leads to the formation of a sufficiently inhomogeneous spatial distribution of
balls whereas for the same system but without trapping, $\triangle=1$, the
balls are actually uniformly distributed over the space region. In the velocity
space the trap effect causes the balls to spread over a much wider domain.

\begin{figure}
\begin{center}
  \includegraphics[width=55mm]{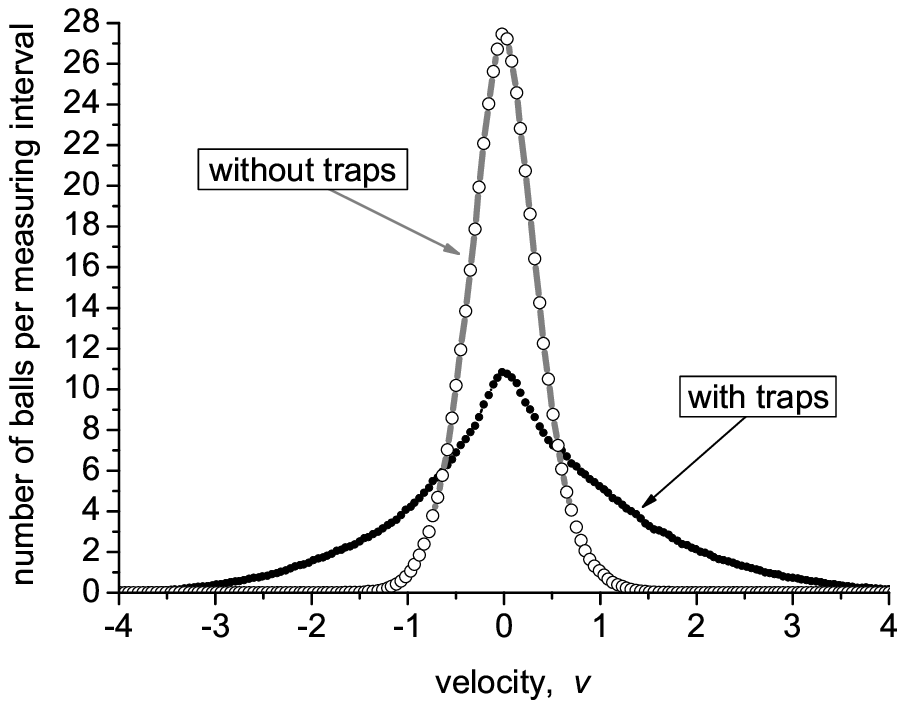}\quad
  \includegraphics[width=55mm]{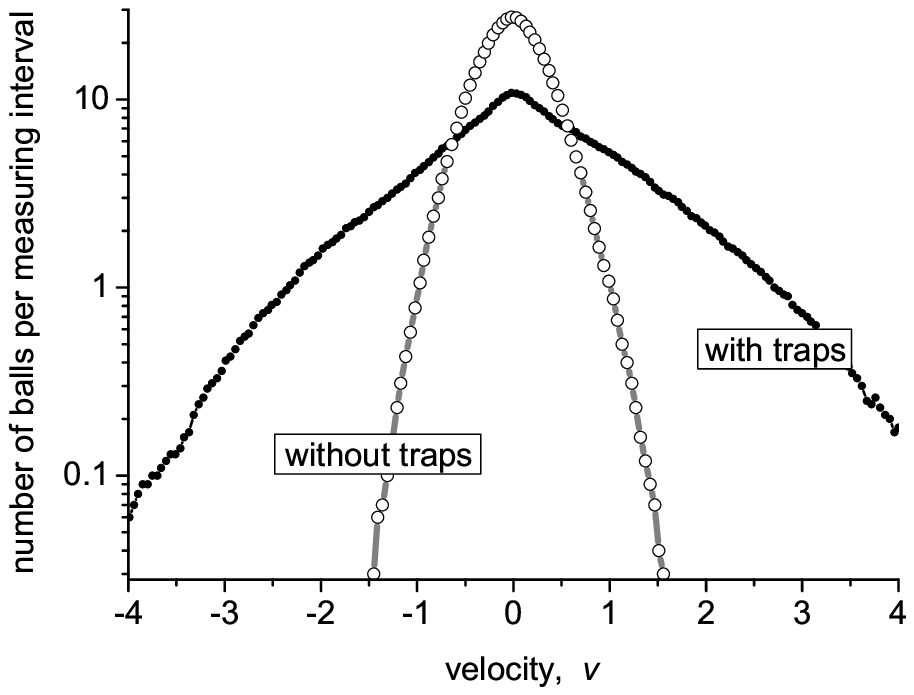}
\end{center}
  \caption{The time averaged velocity distribution presented in the same
  way as in Fig.~\protect\ref{FS1}. Here, however, the individual length of the
  measuring interval was decreased so their total number was 250. The
  parameters used are $\triangle=0.1$, $\sigma =0.1$ and
  $\epsilon = 0.1$, the averaging time interval is 500 unit.
  The left and right windows depict the same data in linear
  and logarithmic scales, respectively.}\label{FS2}
\end{figure}

To make the trap effect in the velocity space more clear we averaged the
velocity distribution over 500 time units. The result is presented in
Fig.~\ref{FS2}. As seen, the strong trap effect causes a nonanalytic behavior
of the velocity distribution at zero value, it takes the form of a cusp.

\begin{figure}
\begin{center}
  \includegraphics[width=80mm]{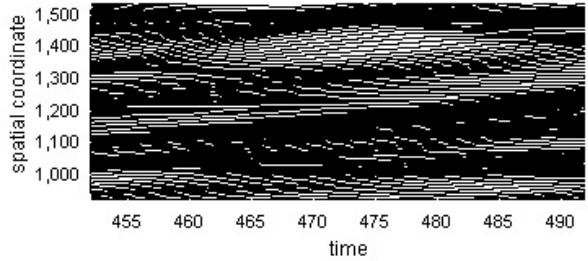}
\end{center}
  \caption{A fragment of the system time pattern. Black regions correspond to
  increased number of balls located in them.}\label{FS3}
\end{figure}

Finally, Fig.~\ref{FS3} illustrates the time pattern of the system dynamics. It
is clear that the shown spatial structure is characterized by a long life time,
because, we recall, the period of individual ball oscillations without traps is
about $2\pi$.

\section{Conclusion}

A new type of noise-induced phase transitions in systems of elements with
motivated behavior is considered. Dynamics of such systems, we think, should
exhibit a number of anomalies due to dynamical traps. The dynamical traps form
a low dimensional region in the phase space where the kinetic coefficients
become sufficiently small and, as a result, the system spends a long time in
it. The cause of the dynamical traps is, e.g., the inability of people or
animals to control all the system governing parameters. So they have to focus
the main attention on a few crucial ones and the intensity of their activity
decreases when the system attains a local quasi-equilibrium with respect to
these parameters.

To illustrate this effect a simple oscillatory system $\{x,v=\dot{x}\}$ is
studied when the trap region is located in the vicinity of the $x$-axis and
without noise the stationary point $\{x=0,v=0\}$ is absolutely stable. For this
system as shown numerically an additive white noise can cause the phase-space
density to take a bimodal shape. For the chain of such oscillators the trap
effect gives rise to a substantially nonuniform spatial distribution and leads
to nonanalytic behavior of the velocity distribution near zero value.

It should be underlined that a possible phase state that could be ascribed in
this case to a maximum of the distribution function in the phase space does not
match any stationary point of the  ``regular'' or ``random'' forces.

\section*{Acknowledgements}
These investigations were supported in part by RFBR~Grants 01-01-00389,
02-02-16537, and Russian Program ``Integration'', Project~B0056.

\end{document}